\documentclass[a4paper,twocolumn]{article}
\usepackage[dvips]{graphicx}
\begin{document}

\title{Information and noise in photon entanglement}

\author{Holger F. Hofmann\\Department of Physics, Faculty of Science, 
University of Tokyo\\7-3-1 Hongo, Bunkyo-ku, Tokyo113-0033, Japan\\
Tel:03-5481-4228, Fax:03-5481-4240\\ 
e-mail: hofmann@femto.phys.s.u-tokyo.ac.jp}

\date{}

\maketitle

\begin{abstract}
By using finite resolution measurements it is possible to simultaneously 
obtain noisy information on two non-commuting polarization components
of a single photon. This method can be applied to a pair of entangled
photons with polarization statistics that violate Bell's inequalities.
The theoretically predicted results show that the non-classical nature 
of entanglement arises from negative joint probabilities for the 
non-commuting polarization components. These negative probabilities 
allow a "disentanglement" of the statistics, providing new insights 
into the non-classical properties of quantum information.
\end{abstract}

\section{Introduction}
One of the most impressive demonstrations of the specifically 
non-classical features of quantum mechanics is the violation of
Bell's inequalities by entangled photon pairs 
\cite{Bel64,Asp82a,Asp82b,Ou88}.
The violation of Bell's inequalities shows that it is impossible 
to explain the statistical predictions of quantum theory by assigning 
a complete set of polarization components to each photon before the 
measurement. This implies that the measurement results for
a specific polarization direction should not be interpreted as a
general property of the photon. In particular, photons cannot be
classified as either x or y polarized, even though only these two 
outcomes are observed in precise polarization measurements along these 
axes.

In the following, the non-classical correlations of entangled photons
are analyzed by applying finite resolution measurements \cite{Hof00a,
Hof00b,Hof00c}. A measurement setup for the simultaneous measurement 
of non-commuting polarization components is presented in section 
\ref{sec:onephoton}. In section \ref{sec:twophotons}, this measurement
concept is applied to entangled photon pairs. It is shown how
information on all four polarization components responsible for the 
violation of Bell's inequalities can be obtained from a single measurement
setup. The analysis of the measurement statistics shows that the violation of 
Bell's inequalities arises from negative joint probabilities similar to the
ones obtained in the single photon measurement setup. The statistics 
derived from the finite resolution measurement thus allows an 
identification of the local non-classical properties responsible for 
the violation of Bell's inequalities. 

\section{Single photon polarization} 
\label{sec:onephoton}
The polarization of light can be characterized by the Stokes parameters,
defined as the intensity difference between orthogonally polarized modes.
A complete description of polarization requires three Stokes parameters.
All Stokes parameters can then be written as components of this three 
dimensional vector. In terms of the annihilation operators for right and 
left circular polarization, $\hat{a}_R$ and $\hat{a}_L$, the Stokes
parameters read
\begin{eqnarray}
\hat{s}_1 &=& \hspace{0.7cm}
\hat{a}^\dagger_R\hat{a}_L + \hat{a}^\dagger_L\hat{a}_R
\nonumber \\
\hat{s}_2 &=& 
-i\left(\hat{a}^\dagger_R\hat{a}_L - \hat{a}^\dagger_L\hat{a}_R \right)
\nonumber \\ 
\hat{s}_3 &=& \hspace{0.7cm}
\hat{a}^\dagger_R\hat{a}_R - \hat{a}^\dagger_L\hat{a}_L.
\end{eqnarray}
For a single photon, these operators have eigenvalues of $\pm 1$, as
observed in measurements using polarization filters. However, a 
polarization filter is only sensitive to one component of the Stokes 
vector at a time, while completely randomizing the information potentially 
carried by the other two components. This limitation can be overcome by 
applying finite resolution measurements to obtain information on one 
polarization component while limiting the noise introduced in the other 
components. It is then possible to study correlations between the 
non-commuting polarization components of a single photon.

\begin{figure}[t]
\setlength{\unitlength}{0.55pt}
\begin{picture}(420,270)
\put(0,200){\line(1,0){360}}
\put(25,230){\makebox(55,15){ \footnotesize Beam}}
\put(25,215){\makebox(55,15){ \footnotesize Displacer}}
\put(30,175){\framebox(45,35){}}
\put(30,200){\line(3,-1){45}}
\put(75,185){\line(1,0){285}}
\put(120,230){\makebox(60,15){ \footnotesize Polarization}}
\put(120,215){\makebox(60,15){ \footnotesize Rotation}}
\put(150,192.5){\circle{30}}
\put(220,220){\makebox(55,15){ \footnotesize Polarizer}}
\put(225,170){\framebox(45,45){}}
\put(225,215){\line(1,-1){45}}
\put(240,200){\line(0,-1){120}}
\put(255,185){\line(0,-1){105}}
\put(335,245){\makebox(60,15){ \footnotesize Detector}}
\put(335,230){\makebox(60,15){ \footnotesize Array}}
\put(360,160){\framebox(10,65){}}
\put(335,140){\makebox(60,15){ \footnotesize $s_2=+1$}}
\bezier{100}(370,170)(370,175)(390,180)
\bezier{100}(390,180)(402,183)(402,185)
\bezier{100}(370,200)(370,195)(390,190)
\bezier{100}(390,190)(402,187)(402,185)
\bezier{100}(370,185)(370,190)(390,195)
\bezier{100}(390,195)(402,198)(402,200)
\bezier{100}(370,215)(370,210)(390,205)
\bezier{100}(390,205)(402,202)(402,200)
\put(140,75){\makebox(60,15){ \footnotesize Detector}}
\put(140,60){\makebox(60,15){ \footnotesize Array}}
\put(215,70){\framebox(65,10){}}
\put(285,67){\makebox(60,15){ \footnotesize $s_2=-1$}}
\bezier{100}(225,70)(230,70)(235,50)
\bezier{100}(235,50)(238,38)(240,38)
\bezier{100}(255,70)(250,70)(245,50)
\bezier{100}(245,50)(242,38)(240,38)
\bezier{100}(240,70)(245,70)(250,50)
\bezier{100}(250,50)(253,38)(255,38)
\bezier{100}(270,70)(265,70)(260,50)
\bezier{100}(260,50)(257,38)(255,38)

\end{picture}
\vspace{-0.5cm}
\setlength{\unitlength}{1pt}

\caption{\label{branch} Finite resolution measurement of non-commuting
polarization components for a single photon.}
\end{figure}

Figure \ref{branch} illustrates the experimental setup for a finite
resolution measurement of two orthogonal components of the Stokes 
vector, $\hat{s}_1$ and $\hat{s}_2$. A beam displacer is used to couple 
the transversal position of the photon with the polarization component
$\hat{s}_1$. The resolution of this measurement is given by the ratio
of the displacement and the width of the beam. After the measurement
of $\hat{s}_1$, the polarization component $\hat{s}_2$ is measured by 
a $\pi/4$ rotation of the polarization axes and a polarizer. However, the
resolution of the $\hat{s}_2$ measurement is limited by the polarization
noise induced in the beam displacer. The detector arrays record the 
continuous measurement values $s_{1m}$ obtained in the measurement of
$\hat{s}_1$ for the two final measurement values of $\hat{s}_2$.

The finite resolution measurement of $\hat{s}_1$ is described by the 
measurement operator \cite{Hof00a}, 
\begin{equation}
\hat{P}_{\delta s}(s_{1m}) = \left(2\pi \delta\!s^2\right)^{-\frac{1}{4}}
\exp\left(- \frac{(\hat{s}_1-s_{1m})^2}{4\delta\!s^2}\right).
\end{equation}
The probability of a measurement of $s_{1m}$ followed by a measurement of
$s_2$ for an input state $\mid\psi_{\mbox{in}}\rangle$ is then given by
\begin{equation}
P(s_{1m};s_2\!=\!\pm 1) =
|\langle s_2\!=\!\pm 1 \mid \hat{P}_{\delta s}(s_{1m}) 
\mid \psi_{\mbox{in}}\rangle|^2.
\end{equation}
\begin{figure}[t]
\setlength{\unitlength}{0.55pt}
\begin{picture}(420,470)

\put(195,210){\makebox(60,20){$P(s_{1m};s_2=-1)$}}
\put(0,15){\makebox(420,200){\includegraphics[width=6cm]{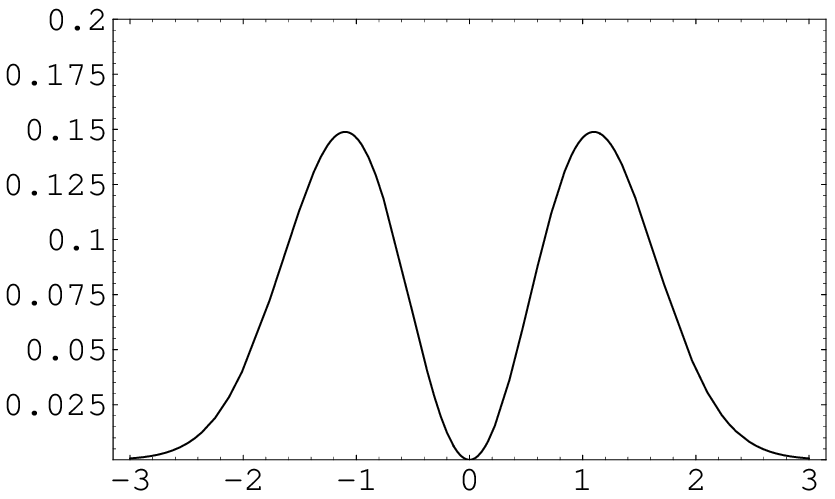}}}
\put(205,0){\makebox(40,20){$s_{1m}$}}

\put(195,450){\makebox(60,20){$P(s_{1m};s_2=+1)$}}
\put(0,255){\makebox(420,200){\includegraphics[width=6cm]{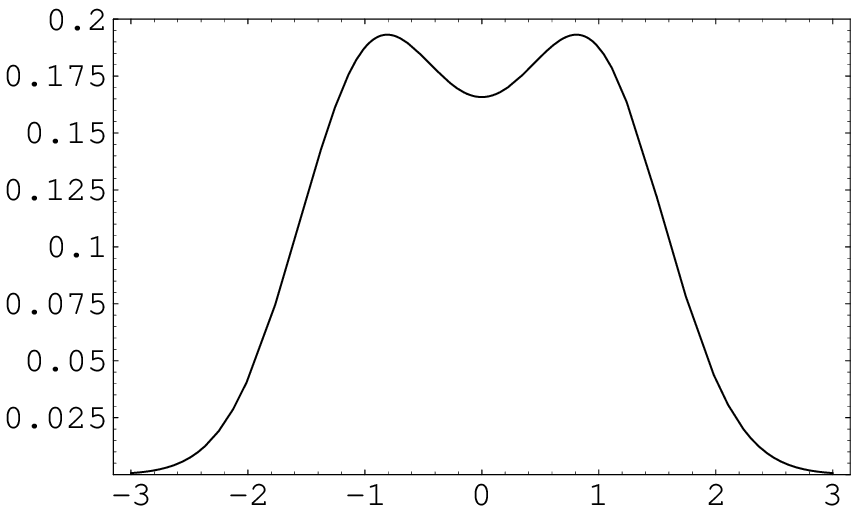}}}
\put(205,240){\makebox(40,20){$s_{1m}$}}

\end{picture}
\setlength{\unitlength}{1pt}

\caption{\label{onephoton} Single photon polarization statistics for 
photons polarized along the $x=y$ direction ($s_2=+1$ eigenstates). 
The resolution is $\delta s=0.6$ for the initial $s_1$ measurement in the
beam displacer.}
\end{figure}
If the input state is in the $+1$ eigenstate of $\hat{s}_2$, the 
measurement statistics are
\begin{eqnarray}
\lefteqn{P(s_{1m};s_2\!=\!+1) =}
\nonumber \\ &&
  \frac{1}{\sqrt{2\pi\delta\!s^2}}
  \exp\left(-\frac{s_{1m}^2+1}{2\delta\!s^2}\right)
  \cosh^2\left(\frac{s_{1m}}{2 \delta\!s^2}\right)
\nonumber \\[0.3cm]
\lefteqn{P(s_{1m};s_2\!=\!-1) =}
\nonumber \\ && 
  \frac{1}{\sqrt{2\pi\delta\!s^2}}
  \exp\left(-\frac{s_{1m}^2+1}{2\delta\!s^2}\right) 
  \sinh^2\left(\frac{s_{1m}}{2 \delta\!s^2}\right),
\end{eqnarray}
as shown in figure \ref{onephoton} for a resolution of $\delta\!s=0.6$. 
The results show that the intuitive assumption that $s_1$ should be 
statistically independent of $s_2$ is wrong even for an eigenstate of
$\hat{s}_2$. Instead, the high values of $s_{1m}$ are clearly correlated 
with ``quantum jumps'' to $\hat{s}_2=-1$. As discussed in \cite{Hof00c},
this implies non-vanishing probability contributions from $s_2=-1$
in the statistics of the $s_2=+1$ eigenstate. 

\begin{table}[bh]
\large
\[
\begin{array}{|c||c|c|c|} \hline
& \multicolumn{3}{|c|}{s_1} \\
s_2 & -1 & 0 & +1 \\
\hline &&& \\[-0.3cm] +1 & \frac{1}{4} & \frac{1}{2} 
& \frac{1}{4} \\[0.3cm]
-1 & \frac{1}{4} & - \frac{1}{2} & \frac{1}{4} 
\\[0.3cm] \hline
\end{array}
\]
\normalsize
\caption{\label{smallstat} Joint probabilities obtained from the
finite resolution measurement setup shown in figure 1 for the $s_2=+1$
eigenstate.} 
\end{table}

If the measurement statistics is interpreted in terms of Gaussian 
contributions with a variance of $\delta\!s^2$ centered around the 
actual values of $s_1$, it appears that, in addition to the quantized
eigenvalue results of $s_1=\pm 1$, results of $s_1=0$ must also be
taken into account. Moreover, the total probability for $s_1=0$ 
remains zero because the negative joint probability of $-1/2$ for 
$s_1=0$ and $s_2=-1$ cancels the positive joint probability of 
$+1/2$ for $s_1=0$ and $s_2=+1$. This negative joint probability also 
explains the coexistence of correlations between $s_1$ and $s_2$ with 
a total probability of zero for $s_2=-1$. The full set of joint 
probabilities obtained for $\delta\!s \to \infty$ is shown in 
table \ref{smallstat}. 

\begin{figure*}[th]
\setlength{\unitlength}{0.9pt}
\hspace{2cm}\begin{picture}(420,420)

\put(20,340){\framebox(60,60){}}
\put(30,380){\makebox(40,15){\large Photon}}
\put(30,365){\makebox(40,15){\large Pair}}
\put(30,350){\makebox(40,15){\large Source}}
\put(85,375){\makebox(15,15){\Large $a$}}
\put(30,320){\makebox(15,15){\Large $b$}}

\put(80,370){\line(1,0){300}}
\put(137,355){\framebox(27,21){}}
\put(137,370){\line(3,-1){27}}
\put(164,361){\line(1,0){216}}
\put(220,365.5){\circle{24}}
\put(287,352){\framebox(27,27){}}
\put(287,379){\line(1,-1){27}}
\put(296,370){\line(0,-1){84}}
\put(305,361){\line(0,-1){75}}
\put(380,346){\framebox(6,39){}}
\put(281,280){\framebox(39,6){}}

\put(383,306){\line(0,-1){110}}
\put(383,196){\line(1,1){15}}
\put(383,196){\line(-1,1){15}}
\put(337,246){\makebox(40,15){\large $s_{1m}(a)$}}
\put(300.5,250){\line(0,-1){54}}
\put(300.5,196){\line(1,1){15}}
\put(300.5,196){\line(-1,1){15}}
\put(254.5,218){\makebox(40,15){\large $s_{1m}(a)$}}

\put(50,340){\line(0,-1){300}}
\put(44,256){\framebox(21,27){}}
\put(50,283){\line(1,-3){9}}
\put(59,256){\line(0,-1){216}}
\put(54.5,200){\circle{24}}
\put(41,106){\framebox(27,27){}}
\put(41,133){\line(1,-1){27}}
\put(50,124){\line(1,0){84}}
\put(59,115){\line(1,0){75}}
\put(35,34){\framebox(39,6){}}
\put(134,100){\framebox(6,39){}}

\put(114,37){\line(1,0){110}}
\put(224,37){\line(-1,1){15}}
\put(224,37){\line(-1,-1){15}}
\put(146,40){\makebox(40,15){\large $s_{1m}(b)$}}
\put(170,119.5){\line(1,0){54}}
\put(224,119.5){\line(-1,1){15}}
\put(224,119.5){\line(-1,-1){15}}
\put(170,122.5){\makebox(40,15){\large $s_{1m}(b)$}}

\put(267,165){\makebox(150,15){\large Coincidence Counts}}

\put(348,2){\framebox(70,70){}}
\put(358,40){\makebox(50,15){\large $s_2(a)=+1$}}
\put(358,19){\makebox(50,15){\large $s_2(b)=+1$}}
\put(348,84.5){\framebox(70,70){}}
\put(358,122.5){\makebox(50,15){\large $s_2(a)=+1$}}
\put(358,101.5){\makebox(50,15){\large $s_2(b)=-1$}}
\put(265.5,2){\framebox(70,70){}}
\put(275.5,40){\makebox(50,15){\large $s_2(a)=-1$}}
\put(275.5,19){\makebox(50,15){\large $s_2(b)=+1$}}
\put(265.5,84.5){\framebox(70,70){}}
\put(275.5,122.5){\makebox(50,15){\large $s_2(a)=-1$}}
\put(275.5,101.5){\makebox(50,15){\large $s_2(b)=-1$}}

\end{picture}
\setlength{\unitlength}{1pt}

\caption{\label{setup} Schematic setup of the coincidence measurement for
entangled photons.}
\end{figure*}

Finite resolution measurements thus reveal that negative joint 
probabilities are an integral part of local quantum statistics.
Since this property of quantum statistics contradicts the assumptions
made about elements of reality in the formulation of Bell's inequalities
\cite{Bel64}, it is possible to explain the violation of Bell's inequalities
by applying the same analysis to the polarization statistics of 
entangled photon pairs.

\begin{figure*}[t]
\hspace{2cm}\setlength{\unitlength}{0.9pt}
\begin{picture}(400,440)
\put(40,420){\makebox(150,20){\large $s_2(a)=-1,\;s_2(b)=+1$}}
\put(40,260){\makebox(150,150){\includegraphics[width=5.2cm]{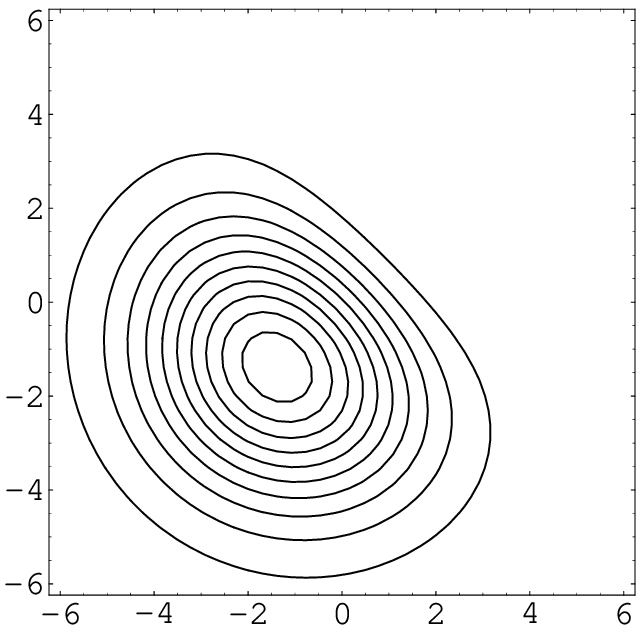}}}
\put(0,335){\makebox(30,10){\large $s_{1m}(b)$}}
\put(105,235){\makebox(40,10){\large $s_{1m}(a)$}}

\put(250,420){\makebox(150,20){\large $s_2(a)=+1,\;s_2(b)=+1$}}
\put(250,260){\makebox(150,150){\includegraphics[width=5.2cm]{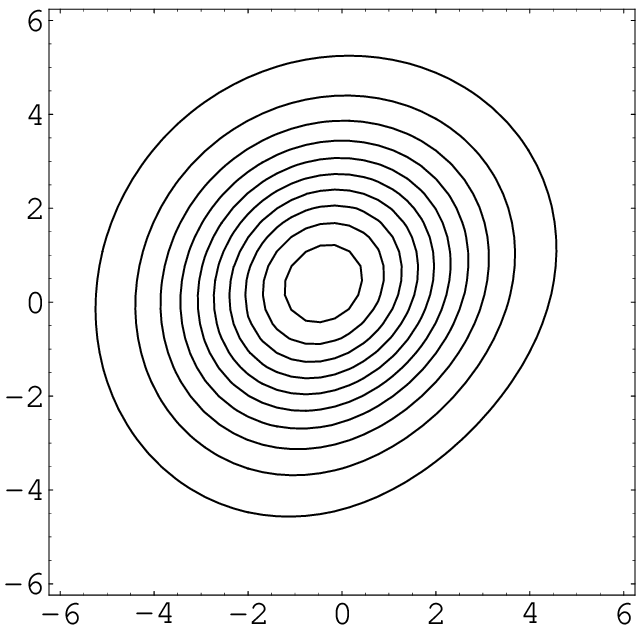}}}
\put(210,335){\makebox(30,10){\large $s_{1m}(b)$}}
\put(315,235){\makebox(40,10){\large $s_{1m}(a)$}}

\put(40,190){\makebox(150,20){\large $s_2(a)=-1,\;s_2(b)=-1$}}
\put(40,30){\makebox(150,150){\includegraphics[width=5.2cm]{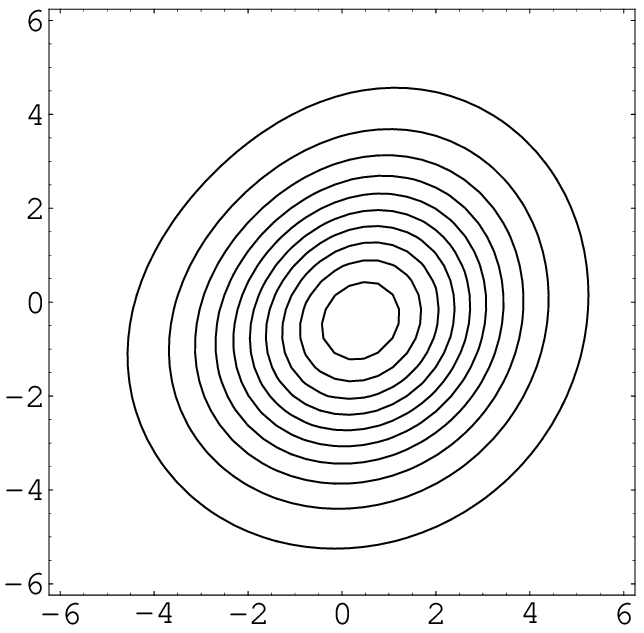}}}
\put(0,105){\makebox(30,10){\large $s_{1m}(b)$}}
\put(105,5){\makebox(40,10){\large $s_{1m}(a)$}}

\put(250,190){\makebox(150,20){\large $s_2(a)=+1,\;s_2(b)=-1$}}
\put(250,30){\makebox(150,150){\includegraphics[width=5.2cm]{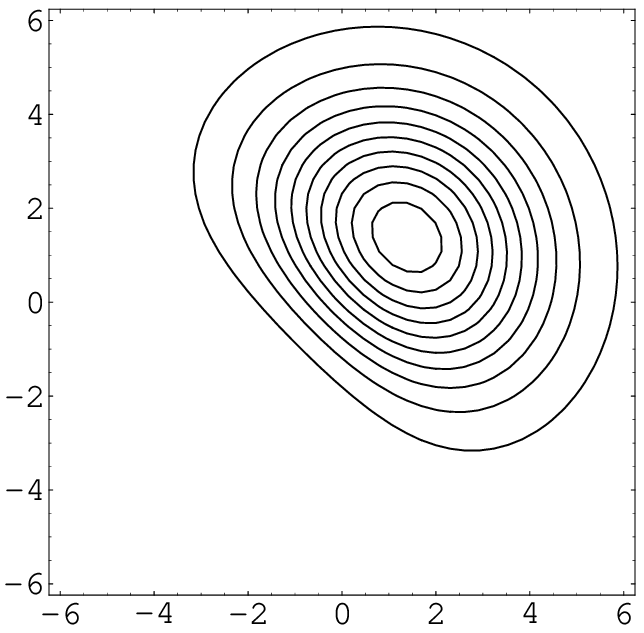}}}
\put(210,105){\makebox(30,10){\large $s_{1m}(b)$}}
\put(315,5){\makebox(40,10){\large $s_{1m}(a)$}}

\end{picture}
\setlength{\unitlength}{1pt}

\caption{\label{stats} Measurement statistics of the coincidence counts
for a resolution of $\delta s = 2$ in the $\hat{s}_1$ measurements.}
\end{figure*}

\section{Polarization statistics of entangled photons}
\label{sec:twophotons}
Figure \ref{setup} shows the setup for a coincidence measurement for
entangled photons. The two branches $a$ and $b$ are set up as illustrated
in figure \ref{branch}. A maximal violation of Bell's inequalities
is obtained for an input state of
\begin{equation}
\mid \psi_{a,b}\rangle = 
\frac{1}{\sqrt{2}}\left(\mid R;L\rangle + 
\exp\left(-i\frac{\pi}{4}\right) \mid L;R\rangle\right),
\end{equation}
where the letters $R$ and $L$ denote eigenstates of right and
left polarization for photon $a$ and photon $b$, respectively. 
This input state is an eigenstate of the correlation
\begin{eqnarray}
\label{eq:K}
\hat{K}&=&\hat{s}_1(a)\hat{s}_1(b)+\hat{s}_2(a)\hat{s}_1(b)
\nonumber \\ &&
-\hat{s}_1(a)\hat{s}_2(b)+\hat{s}_2(a)\hat{s}_2(b)
\end{eqnarray}
with an eigenvalue of $K=2\sqrt{2}$. The maximal value obtained by 
assigning eigenvalues of $\pm 1$ to the operators $\hat{s}_i(a/b)$
in equation (\ref{eq:K}) is $K=2$.
Therefore, $\mid \psi_{a,b}\rangle$ maximally violates the Bell's 
inequality $K\leq 2$.

Figure \ref{stats} shows the measurement statistics obtained for a
resolution of $\delta\!s=2$. At this resolution, the quantum noise
introduced in the measurement of $\hat{s}_1$ is still very low, so
that the original properties of $\hat{s}_2$ are preserved. Therefore, 
the statistics clearly reveal the non-classical features of correlations
between $\hat{s}_1$ and $\hat{s}_2$. In particular, the peaks of the results
obtained for $s_2(a)=-s_2(b)$ are at values of $s_{1m}=\pm \sqrt{2}$,
far beyond the eigenvalue limits of $\pm 1$. Moreover, the peaks are
actually sharper than the resolution of $\delta\!s=2$ would allow in a
classical context. 

\begin{table*}[t]
\large
\[
\begin{array}{|c||c|c|c||c|c|c|} \hline 
&\multicolumn{6}{|c|}{}\\
&\multicolumn{6}{|c|}{\left(s_1(a),s_2(a)\right)}\\[0.3cm]
\left(s_1(b),s_2(b)\right) & \hspace{1.8cm} & \hspace{1.8cm} & \hspace{1.8cm} &
\hspace{1.8cm} & \hspace{1.8cm} & \hspace{1.8cm} \\[-0.3cm]
& (-1,-1) & (0,-1) & (1,-1) & 
(-1,1) & (0,1) & (1,1) \\[0.5cm]
\hline &&&&&&\\
(1,1) & \frac{\sqrt{2}-1}{16\sqrt{2}} & - \frac{1}{8\sqrt{2}} & 
\frac{\sqrt{2}+1}{16\sqrt{2}} & 
\frac{\sqrt{2}-1}{16\sqrt{2}} & \frac{1}{8\sqrt{2}} &
\frac{\sqrt{2}+1}{16\sqrt{2}} \\[0.5cm]
(0,1) & \frac{1}{8\sqrt{2}} & - \frac{1}{4\sqrt{2}} & 
- \frac{1}{8\sqrt{2}} & 
\frac{1}{8\sqrt{2}} & \frac{1}{4\sqrt{2}} &
-\frac{1}{8\sqrt{2}} \\[0.5cm]
(-1,1) & \frac{\sqrt{2}+1}{16\sqrt{2}} & \frac{1}{8\sqrt{2}} & 
\frac{\sqrt{2}-1}{16\sqrt{2}} & 
\frac{\sqrt{2}+1}{16\sqrt{2}} & - \frac{1}{8\sqrt{2}} &
\frac{\sqrt{2}-1}{16\sqrt{2}} \\[0.5cm]
\hline &&&&&&\\
(1,-1) & \frac{\sqrt{2}-1}{16\sqrt{2}} & - \frac{1}{8\sqrt{2}} & 
\frac{\sqrt{2}+1}{16\sqrt{2}} & 
\frac{\sqrt{2}-1}{16\sqrt{2}} & \frac{1}{8\sqrt{2}} &
\frac{\sqrt{2}+1}{16\sqrt{2}} \\[0.5cm]
(0,-1) & -\frac{1}{8\sqrt{2}} & \frac{1}{4\sqrt{2}} & 
\frac{1}{8\sqrt{2}} & 
- \frac{1}{8\sqrt{2}} & -\frac{1}{4\sqrt{2}} &
\frac{1}{8\sqrt{2}} \\[0.5cm]
(-1,-1) & \frac{\sqrt{2}+1}{16\sqrt{2}} & \frac{1}{8\sqrt{2}} & 
\frac{\sqrt{2}-1}{16\sqrt{2}} & 
\frac{\sqrt{2}+1}{16\sqrt{2}} & - \frac{1}{8\sqrt{2}} &
\frac{\sqrt{2}-1}{16\sqrt{2}} \\[0.5cm]
\hline
\end{array}
\]
\caption{\label{negprop} Joint probabilities for the non-commuting
spin components. Note that the total probability for $s_1$ values of
zero is always zero.}
\end{table*}

\begin{table*}[t]
\large
\[
\begin{array}{|c||c|c|c||c|c|c|} \hline 
&\multicolumn{6}{|c|}{}\\
&\multicolumn{6}{|c|}{\left(s_1(a),s_2(a)\right)}\\[0.3cm]
\left(s_1(b)\!,s_2(b)\right) & 
\hspace{1.8cm} & \hspace{1.8cm} & \hspace{1.8cm} &
\hspace{1.8cm} & \hspace{1.8cm} & \hspace{1.8cm} \\[-0.3cm]
& (-1,-1) & (0,-1) & (1,-1) & 
(-1,1) & (0,1) & (1,1) \\[0.5cm]
\hline &&&&&&\\
(1,1)  & -2 & -2 & -2 & +2 & +2 & +2 \\[0.5cm]
(0,1)  &  0 & -1 & -2 & +2 & +1 &  0 \\[0.5cm]
(-1,1) & +2 &  0 & -2 & +2 &  0 & -2 \\[0.5cm]
\hline &&&&&&\\
(1,-1) & -2 &  0 & +2 & -2 &  0 & +2 \\[0.5cm]
(0,-1) &  0 & +1 & +2 & -2 & -1 &  0 \\[0.5cm]
(-1,-1)& +2 & +2 & +2 & -2 & -2 & -2 \\[0.5cm]
\hline
\end{array}
\]
\caption{\label{Kvalue} Values of the correlation K for different 
joint values of the polarization components.}

\end{table*}

As in the case of a single photon, the statistics may be interpreted as
a sum of Gaussian contributions with a variance of $\delta\!s^2$ centered 
around the actual values of $s_1$. The non-classical features then arise
from the negative joint probabilities at $s_1(a)=0$ and/or $s_1(b)=0$.
The sharpness and the shift of the peaks at $s_2(a)=-s_2(b)$ are
explained by the negative probability at $s_1(a)=s_1(b)=0$. Table 
\ref{negprop} shows the full set of joint probabilities obtained from the
measurement results for $\delta\!s \to \infty$. 

As shown in table \ref{Kvalue}, each of the 36 measurement results in 
table \ref{negprop} corresponds to a well defined value of $K$. 
In accordance with the probability maxima in figure \ref{stats}, the
values of $K=+2$ are found at $s_1(a)=+1$ or $s_1(b)=-1$ for
$s_2(a)=s_2(b)=-1$, at $s_1(a)=-1$ or $s_1(b)=+1$ for $s_2(a)=s_2(b)=+1$, 
at $s_1(a)=-1$ and $s_1(b)=-1$ for $s_2(a)=-1$ and $s_2(b)=+1$, 
and at $s_1(a)=+1$ and $s_1(b)=+1$ for $s_2(a)=+1$ and $s_2(b)=-1$.
The broadness of the peaks in the measurement statistics observed for
$s_2(a)=s_2(b)$ in figure \ref{stats} is explained by the positive
probability contribution for $K=+1$ at $s_1(a)=s_1(b)=0$. The steep
slopes of the peaks for $s_2(a)=-s_2(b)$ is likewise explained by the
negative probability contribution for $K=-1$ at $s_1(a)=s_1(b)=0$.
The regions of low probability in figure \ref{stats} are explained by
the near cancellation of negative and positive probabilities for
values of $K=-2$ at $s_1(a)=+1$ or $s_1(b)=+1$ for
$s_2(a)=-1$ and $s_2(b)=+1$, at $s_1(a)=-1$ or $s_1(b)=-1$ for 
$s_2(a)=+1$ and $s_2(b)=-1$, at $s_1(a)=-1$ and $s_1(b)=+1$ for 
$s_2(a)=s_2(b)=-1$, and at $s_1(a)=+1$ and $s_1(b)=-1$ for 
$s_2(a)=s_2(b)=+1$. The total probability distribution of K values
then reads
\begin{eqnarray}
P(K=2) = 103.1 \% && P(K=-2) = -3.1 \%
\nonumber \\
P(K=1) = 35.4 \% && P(K=-1) = -35.4 \%
\nonumber \\
P(K=0) = 0 \%. 
\end{eqnarray}
The violation of Bell's inequalities is therefore the result of 
negative joint probabilities for the non-commuting polarization
components of the entangled photon pair.

\section{Elements of reality and negative probabilities}
\label{sec:elements}
The formulation of Bell's inequalities is based on the assumption
that the operator variables can be represented by their eigenvalues.
This assumption reflects the definition of elements of reality
given in the famous paper by Einstein, Podolsky and Rosen (EPR) \cite{EPR}:
``{\it If, without in any way disturbing a system, we can predict
with certainty (i.e., with probability equal to unity) the value of a 
physical quantity, then there exists an element of physical reality
corresponding to this physical quantity.}'' However, this argument 
breaks down if the statistics include negative probabilities. 
If one value of a physical quantity has a probability equal unity, 
it is still possible that another value of the same property has 
positive and negative joint probabilities. In particular, the discussion 
of single photon polarization in section \ref{sec:onephoton} revealed
the presence of contributions from $s_1=0$, even though no eigenvalue 
of $\hat{s}_1$ corresponds to this result. While it is possible to 
predict with certainty that no precise measurement of $\hat{s}_1$ can
produce this result, this certainty does not apply to
the result of finite resolution measurements. While the total probability
for $s_1=0$ is always zero, the joint probabilities shown in tables
\ref{smallstat} and \ref{negprop} are not. Negative probabilities thus 
introduce a measurement dependent ambiguity into the selection of elements 
of reality that contradicts the assumptions of Bell's inequalities. 

Note that negative probabilities cause no conceptual problems as long
as the uncertainty principle applies to all measurements. Indeed, the 
uncertainty principle can be interpreted as a consequence of negative
joint probabilities since it must be impossible to isolate an event
associated with a negative probability. Uncertainty guarantees
that negative probabilities are always ``covered up'' by quantum noise
in the measurement process. Effectively, actual measurement results
can only be associated with a region of phase space sufficiently large
to include more positive than negative probability contributions.

\section{Conclusions}
\label{sec:concl}
Finite resolution measurements of single photon polarization allow
simultaneous measurements of non-commuting Stokes parameter components.
By applying this type of measurement to entangled photon pairs,
details of the violation of Bell's inequalities can be obtained in
a single measurement setup. It is possible to represent the statistics 
of the photon pair polarization in a table of 36 joint probabilities for 
the non-commuting polarization components. Non-classical features arise 
from the negative probabilities at values of $s_1=0$. These features not 
only explain the violation of Bell's inequalities, but also establish a 
connection between entanglement and the non-classical properties of 
individual quantum systems.

\section*{Acknowledgements}
I would like to acknowledge support from the Japanese 
Society for the Promotion of Science, JSPS.




\begin{thebibliography}{5}
\bibitem{Bel64} J.S. Bell, Physics {\bf 1}, 195 (1964).

\bibitem{Asp82a}
A. Aspect, P. Grangier, and G. Roger, Phys. Rev. Lett. {\bf 49}, 91 (1982).

\bibitem{Asp82b}
A. Aspect, J. Dalibard, and G. Roger, Phys. Rev. Lett. {\bf 49}, 1804 (1982).

\bibitem{Ou88}
Z. Y. Ou and L. Mandel, Phys. Rev. Lett. {\bf 61}, 50 (1988).

\bibitem{Hof00a}
H.F. Hofmann, Phys.Rev. A {\bf 62}, 
022103 (2000).

\bibitem{Hof00b}
H.F. Hofmann, Phys.Rev. A {\bf 61}, 033815 (2000).

\bibitem{Hof00c}
H.F. Hofmann, T. Kobayashi, and A. Furusawa, 
Phys.Rev. A {\bf 62}, 013806 (2000).

\bibitem{EPR}
A. Einstein, B. Podolsky, and N. Rosen, Phys. Rev. {\bf 47}, 777 (1935).

\end{thebibliography}
\end{document}